\documentclass[conference]{IEEEtran}
\ifCLASSINFOpdf
\else
\fi
\begin{document}
%
\title{Calculations for electron-impact ionization of beryllium in the method of interacting configurations in the complex number representation}


\author{\IEEEauthorblockN{\textbf{V.M. Simulik, T.M. Zajac*, R.V. Tymchyk}}

\\
\IEEEauthorblockA{Institute of Electron Physics, National Academy of Sciences of Ukraine,\\
21 Universitetska Str., Uzhgorod, 88000, Ukraine,
\\ Department of Electronic Systems*, Uzhgorod National University, \\ 13 Kapitulna Str., Uzhgorod, 88000, Ukraine\\
Email: vsimulik@gmail.com}
}


%


\maketitle

\begin{abstract}
The beginning of the application of the method of interacting configurations in the complex number representation to the compound atomic systems has been presented. The spectroscopic characteristics of the Be atom in the problem of the electron-impact ionization of this atom are investigated. The energies and the widths of the lowest $^{1}S,\,^{1}P,\,^{1}D,$ and $^{1}F$𝐹autoionizing states of Be atom are calculated.
\end{abstract}


%
\IEEEpeerreviewmaketitle

\pagestyle{plain}

\section{Introduction}
The method of interacting configurations in the complex number representation (ICCNR method) is applied to the calculation of spectroscopic characteristics of autoionizing states (AIS) of Be atom in the problem of the electron-impact ionization of this atom. In particular, the energies and the widths of the lowest ($^{1}S,\,^{1}P,\,^{1}D,$ and $^{1}F$) AIS of Be atom are calculated. 

The exact quantum mechanical methods are welcome here because the experimental investigations of beryllium are complicated due to its chemical properties. Furthermore, the atomic shell of Be atom is not simple and is not very complex. Therefore, this atom is enough suitable object for the beginning of the application of the ICCNR method to the compound atomic systems investigations.

Investigations of autoionization phenomena in the frame-work of the problems dealing with the ionization and the electron scattering by atoms and ions were separated in the last decades into an independent branch of theoretical atomic physics. The scientific interest to the description of the processes of excitation and decay of quasistationary states is associated with a necessity to specify the parameters of elementary processes, which are used in theoretical estimations and calculations in plasma physics, laser spectroscopy, solid state physics, and crystallography, at the development of technological methods of isotope separation at the atomic level, the designing of coherent ultra-violet and x-ray radiation generators, as well as in other physical domains.

The results of experimental researches concerning the AIS located between the first and second ionization thresholds for helium and helium-like ions were qualitatively explained on the basis of the theory of isolated Fano resonance and in the diagonalization approximation. The appearance of new experimental data on resonance structures in partial cross-sections of helium photoionization above the threshold of excited ion formation (more exactly, in the interval between the second and third thresholds, to which the AIS energies converge in the atomic ionization problem) brought about a number of theoretical issues dealing, first of all, with the description of the interaction of a considerable number of overlapping quasistationary states, which decay through several open channels. Theoretical calculations and the analysis of resonance structures decaying into several states of a residual ion should be carried out, in the general case, with regard for all interconfiguration interactions.

One of the first theoretical methods that made it possible to obtain results coinciding with experimental data was the method of overlapping configurations or the method of interacting configurations. In the terminology adopted in this work, this formalism is called the method of interacting configurations in the real number representation (see Section 3). An important step of the theory became the ICCNR method. This method was developed in works [1--3] and successfully applied to the description of the quasistationary states of helium formed at its electron ionization in the energy interval above the threshold of excited ion formation.

At the modern stage in the development of this method, a principal advantage is its application to the calculation of ionization processes in more complicated atomic structures, see, e. g., [4]. As one can see in the literature [5--15], beryllium atom turns out to be the promising objects for researches.

Here some results from [5--10] are compared with the calculations on the basis of ICCNR method. Further analysis of literature shows that beryllium atom until today is the interesting experimental and theoretical problem, see, e. g., [11--15]. The authors of [11--15] prefer to approbate their experimental and theoretical methods in the investigations of processes in the beryllium atomic shell.     

\section{General description of the method}

The ICCNR method is used here to calculate the energies and the widths of quasistationary states in the problem of electron impact ionization of Be atom. In this section, the fundamentals and the formalism of the method are briefly described.

Since we deal with the quantum-mechanical three body problem (see the reaction (1) below) let us consider briefly the main methods for studying the three-particle quantum-mechanical scattering phenomena.

The first well-defined mathematical method is based on Faddeev equations, see, e. g., [16--18]. Indeed, some times the three body problem in quantum mechanics can be solved with the help of Faddeev equations. Nevertheless, this method is applied rarely for the calculations in atomic problems. The reason is in the long-range property of Coulomb potential and, as a consequence, in the slow conver-gence of the corresponding series. Therefore, usually the method is applied for the simplest atomic systems [16--18]. As an example, the method was applied for the (e, H) system, in which the hydrogen negative ion bound state and the lowest members of the resonances in both the singlet and the triplet $J=0$ series were calculated [16]. Further, in [17] the AIS of He atom were calculated. The resonance ionization by fast protons was studied, which was possible for the short-range nuclear potential. In [18] the scattering of electrons and positrons on hydrogen atom below the n=2 threshold was investigated. Thus, the Faddeev equations are not applied for the complex atoms AIS calculations.

On the other hand both the Faddeev and ICCNR methods are based on the Lippmann-Schwinger-“type” equations. Indeed, the Faddeev formalism is started from the indication of the absence of uniqueness in the Lippmann-Schwinger solutions. This fact indicates some similarity between methods. However, the difference appears in T-operator analysis. Moreover, we do not appeal to three particle potential.

Another well-defined mathematical method, which has some similarity to Faddeev approach, is the method of hyperspherical functions (hyperspherical coordinates). The success of this method is evident from [19, 20]. However, this method is working good only for few body systems (here with Coulomb interaction too) and is not applied for the complex atoms AIS investigations as well.

Note that the R-matrix method is the most spread in the complex atoms AIS investigations [21, 22]. This method is widely applied for the scattering of photons and electrons on the complex atoms both in ordinary [21] and B-spline [22] formalism.

The ICCNR method is a well-defined quantum-mechanical method for the calculation of parameters of atomic systems. This method is a development and a generalization of the known method of interacting configurations in the real number representation. It has some advantages in comparison with the standard method of interacting configurations in the real number representation and other calculation methods for the energies and widths of quasistationary atomic states. First, this is a capability of finding not only the energies, but also the widths of quasistationary states. Second, there are new possibilities for the resonance identification. The ICCNR method makes it possible, on the basis of the results of calculations, to estimate the contribution of each resonance state to the cross-section of the process and, if the resonance approximation is applicable, to introduce a set of parameters that determine the energies and the widths of quasistationary states, as well as the contours of resonance lines in the ionization cross-sections. This approach also enables the applicability of approximate methods to the estimation of cross-sections in specific problems to be studied and the limits of their validity to be determined. Those advantages make it possible to successfully apply the ICCNR method not only to scattering processes, but also to much more complicated processes of atomic ionization by electrons.

The goal of our research is to illustrate the capabilities of the ICCNR method in the determination of spectroscopic characteristics of complicated atoms. Quasistationary states were studied in such multielectron atomic systems as Be atom [4]. The capabilities of the method were illustrated briefly by the example of the atomic ionization by the electron impact [4], which are challenging for researches. The analysis of the loss spectrum of ejected electrons made it possible to compare indirectly the obtained results with the results of studies of the scattering problem. The results were reported at the international conference [4]. This preprint is the expanded presentation of [4].

\section{Formalism of the method}

Let us recall the foundations of the ICCNR method for the study of the processes of atomic ionization by the electron impact. Consider the equation of the examined reaction
\begin{equation}
\label{eq1}
A(n_{0}L_{0}S_{0})+e^{-}(\overrightarrow{k}_{0}) \rightarrow A^{+}(nl_{1})+e^{-}(\overrightarrow{k}_{1})+e^{-}(\overrightarrow{k}),
\end{equation}

\noindent where $\overrightarrow{k}_{0},\overrightarrow{k}_{1},\overrightarrow{k}$ are the momenta of the incident, ejected, and scattered electrons, respectively. Then the generalized oscillator strength of the transition for the incident electron in the Born approximation is given by
\begin{equation}
\label{eq2}
\frac{df_{nl_{1}}}{dE}(Q)=\frac{E}{Q^{2}} \sum_{lL} |\langle nL_{1}El|\sum_{j=1}^{n}\exp(i\overrightarrow{Q}\overrightarrow{r_{i}})|n_{0}L_{0}S_{0}\rangle |^{2}.
\end{equation}

\noindent In this formula  $E=k_{0}^{2}-k^{2}$ is the energy loss, $\overrightarrow{Q}=\overrightarrow{k_{0}}- \overrightarrow{k}$ is the transmitted momentum, and $|nl_{1}El:LS_{0}\rangle$ is the wave function of an atom with total momentum $L$ and spin $S_{0}$ provided that an electron with momentum $l$ and energy $E$ is in the field of ion $A^{+}$, whose electron has the quantum numbers $|nl_{1}\rangle$. The function of the atomic ground state is given by $|n_{0}L_{0}S_{0}\rangle$.

Note that process (1) is a much more complicated physical phenomenon in comparison with the electron scattering by an atom. Exact theoretical calculations of such processes constitute a problem for modern theoretical physics. Therefore, the consideration of this problem for multielectron atoms in the framework of the ICCNR method is an important and challenging scientific step.

The choice of the wave function for the ground state is dictated by a desirable accuracy of the final results of cal-culations. In the case of two-electron systems, this is a mul-tiparametric Hylleraas-type wave function [23], and, in the case of Be atom, this is, as a rule, a Hartree–Fock wave function obtained in the multiconfiguration approximation [24]. The system of equations in the ICCNR method has the following form:	
\begin{equation}
\label{eq3}
(E_{n}-E)a^{Ei}_{\lambda n}+ \sum_{\lambda'}\int^{\infty}_{0}b^{Ei}_{\lambda \lambda'}(E')V_{n\lambda'}(E')dE',
\end{equation}
$$\sum_{m}a^{Ei}_{\lambda m}V^{\ast}_{m\lambda'}(E')+(E'-E)b^{Ei}_{\lambda \lambda'}(E')=0.$$
\noindent The multipliers $a^{Ei}_{\lambda m}$ and $b^{Ei}_{\lambda \lambda'}(E')$ are the coefficients of expansion of the wave function $\Psi^{E}_{\lambda}(\overrightarrow{r_{1}},\overrightarrow{r_{2}})$ in the basis
\begin{equation}
\label{eq4}
\Psi^{E}_{\lambda}(\overrightarrow{r_{1}},\overrightarrow{r_{2}})=\sum_{m}a^{Ei}_{\lambda m}|m\rangle+\sum_{\lambda'}\int^{\infty}_{0}b^{Ei}_{\lambda \lambda'}(E')|\lambda'E'\rangle dE'.
\end{equation}

\noindent The basis wave functions satisfy the conditions
\begin{equation}
\label{eq5}
\langle m|\widehat{H}|n\rangle=E_{n}\delta_{nm}, \, \langle\lambda'E'|\widehat{H}|\lambda E\rangle=E\delta_{\lambda\lambda'}\delta(E-E'),
\end{equation}
\noindent where $\widehat{H}$ is the total Hamiltonian of the system.

The formal solution for the multiplier $b^{Ei}_{\lambda \lambda'}(E')$ is given by 
\begin{equation}
\label{eq6}
 b^{Ei}_{\lambda \lambda'}(E')=P\frac{\sum_{m}a^{Ei}_{\lambda m}V_{m\lambda}(E)}{E-E'}+[A_{\lambda\lambda'}
\end{equation}
$$\pm i\pi\sum_{m}a^{Ei}_{\lambda m}V_{m\lambda'}(E)]\delta(E-E'),$$
\noindent where $V_{m\lambda}(E)=\langle m|\widehat{H}|\lambda E\rangle$. The matrix $A_{\lambda\lambda'}$ depends on the asymptotic properties of the basis functions $|\lambda E\rangle$. Substitution of Eq. (6) into Eq. (3) transforms the system of equations obtained in the ICCNR method into a system of linear algebraic equations for the coefficients $a^{Ei}_{\lambda m}$:
\begin{equation}
\label{eq7}
(E_{n}-E)a^{Ei}_{\lambda n}+\sum_{m}[F_{nm}(E)-i\gamma_{nm}(E)]a^{Ei}_{\lambda m}
\end{equation}
$$=-\sum_{\lambda'}A_{\lambda\lambda'}V_{\lambda'n}(E),$$
\noindent The latter can be expressed in terms of eigenvectors and eigenvalues of the complex matrix
\begin{equation}
\label{eq8}
W_{nm}(E)=E_{n}\delta_{nm}+F_{nm}(E)-i\gamma_{nm}(E),
\end{equation}
\noindent where
\begin{equation}
\label{eq9}
\gamma_{nm}(E)=\pi\sum_{\lambda}V_{n\lambda}(E)V_{\lambda m}(E);
\end{equation}
$$F_{nm}(E)=\frac{1}{\pi}\int^{\infty}_{0}\frac{\gamma_{nm}(E)}{E-E'}dE'.$$

The analysis of formulas (8) and (9) allows one to compare various approximations, which can be done in the ICCNR method. One can see that, in the framework of this method, the following approximations are possible:

 1) the method of interacting configurations in the real number representation; this approximation corresponds to the neglect of complex components $i\gamma_{nm}(E)$ in matrix (8);

2) the diagonalization approximation in the real number representation consists in that the sum of all non-diagonal members $F_{nm}(E)-i\gamma_{nm}(E)$ in the matrix $W_{nm}(E)$ is neglected;

3) the diagonalization approximation involving the tran-sitions outside the energy surface (or the diagonalization approximation in the complex number representation) arises if the term $F_{nm}(E)$ is neglected in calculations.

The account for all members in matrix (8) is, in essence, the ICCNR method, the advantages of which over the indicated approximations are obvious.

After determining the eigenvectors and eigenvalues of the matrix $W_{nm}(E)$, we can calculate the energies and widths of quasistationary states that are located above the threshold of excited ion formation [1, 2]. The partial amplitudes of the resonance ionization can be determined as follows:
\begin{equation}
\label{eq10}
T_{|0\rangle \rightarrow|\lambda E\rangle}(E)=t^{dir}_{\lambda}(E)+\sum_{m}\frac{H_{m\lambda(E)}}{\varepsilon_{m}(E)+1}.
\end{equation}

\noindent The quantities in formula (10) are defined by the relations
\begin{equation}
\label{eq11}
t^{dir}_{\lambda}(E)=\sqrt{C(E)}\langle\lambda E|\hat{t}|0\rangle,
\end{equation}
\begin{equation}
\label{eq12}
H_{m\lambda}(E)=2\widetilde{V}_{m\lambda}(E)[t_{m}(E)-i\tau_{m}(E)]\Gamma^{-1}_{m}(E),
\end{equation}
\noindent where
\begin{equation}
\label{eq13}
t_{m}(E)=\sqrt{C(E)}\langle \widetilde{F}^{E}_{m}|\hat{t}|0\rangle, \quad \tau_{m}(E)=\sqrt{C(E)}\langle \chi^{E}_{m}|\hat{t}|0\rangle.
\end{equation}
\noindent Hence, the expressions for the cross-sections become parametrized
\begin{equation}
\label{eq14}
\sigma_{\lambda}(E)=\sigma_{\lambda}^{dir}(E)+\sum_{m}\frac{\Gamma_{m}(E)P_{m\lambda}(E)+\varepsilon_{m}(E)
Q_{m\lambda}(E)}{\varepsilon^{2}_{m}(E)+1}.
\end{equation}

The real functions $P_{m\lambda}(E)$ and $Q_{m\lambda}(E)$ of the total energy $E$ are the doubled real and imaginary, respectively, parts of the complex function $N_{m\lambda}(E)$, which looks like
\begin{equation}
\label{eq15}
N_{\alpha m}(E)\nonumber=\sum_{\lambda\varepsilon\alpha}H_{m\lambda}(E)(t^{dir}_{\lambda}(E)+\sum_{n}\frac{H_{m\lambda}(E)}{\varepsilon_{n}(E)-\varepsilon_{m}(E)+2i})^{\ast}.
\end{equation}
Hence, the resonance ionization cross-section is determined by a collection of the following functions of the total energy $E$: $\sigma_{\lambda}^{dir}(E)$, $N_{\alpha m}(E)$, $\varepsilon_{m}(E)$, and $\Gamma_{m}(E)$. See more details about the formalism of the method (for two electron systems) in the article [25]. 

\section{The results of the calculations}

Here the electron-impact ionization of the Be atom in the interval of AIS excitation is considered.

In work [4], using the ICCNR method, the research of the ionization of a Be atom by the electron impact in the AIS excitation interval was started, and the spectra of energy loss were analyzed. The photoionization of this atom was studied as well. The AIS that arise at that can be compared with the AIS that are formed in the problem of electron scattering at the corresponding ion. In calculations, the Coulomb wave functions were used as basis configurations. For every term, up to 25 basis configurations were taken into account.

Table 1 contains the results of our calculations for the energies and the widths of the lowest AIS of a Be atom ($^{1}S,\,^{1}P,\,^{1}D,$ and $^{1}F$) obtained in the problem of the ionization of this atom by the electron impact with the use of the ICCNR method [4]. The results are compared with the energies and the widths of AIS obtained in the problem of electron scattering by a Be$^{+}$ ion in work [8]. Therefore, this comparison is indirect. In addition, in Table 2, the energies of $^{1}P$ states, which are located between the first and second ionization thresholds of a beryllium atom, are compared with the results of calculations obtained by other authors [5--10].

In the literature, there are no similar results obtained on the basis of exact computational methods, in particular, on the basis of the method of interacting configurations and, the more so, on the basis of the ICCNR one. The comparison with the results of calculations of corresponding autoionizing state energies in the problem of electron scattering by Be$^{+}$ ions performed in work [8] in the diagonalization approximation (see Table 1) is indirect, because it deals with a different object in a different problem. Nevertheless, it really evidences the reliability of the results obtained here.

\begin{table}[h] \caption{Energies and widths of the lowest AIS ($^{1}S,\,^{1}P,\,^{1}D,$ and $^{1}F$) of a beryllium atom obtained in the ICCNR approximation in the problem of the electron-impact ionization of an atom. In the paper [8], the energies of AIS were calculated in the diagonalization approximation in the framework of the problem of electron scattering by a Be$^{+}$ ion}

\tabcolsep4.2pt

\begin{center}

\begin{tabular}{|c|c|c|c|c|}
\hline
\rule{0pt}{5mm} $^{1}S$  & E, eV & $\Gamma$, eV & E, eV [8]& $\Gamma$ eV [8]\\
\hline
$3s^{2}$ & 16.42 & 0.0803  & 16.40 & 0.0818 \\
\hline
$3p^{2}$ & 18.65 & 0.0110  & 18.57 & 0.0116 \\
\hline
$3s4s$   & 18.82 & 0.0351  & 18.74 & 0.0358 \\
\hline
$3s5s$   & 19.48 & 0.0163  & 19.45 & 0.0167 \\
\hline
$3s6s$   & 19.77 & 0.00869 & 19.75 & 0.00884 \\
\hline
$3s7s$   & 19.96 & 0.00518 & 19.92 & 0.00527 \\
\hline
$^{1}P$  & E, eV & $\Gamma$ eV & E, eV [8]& $\Gamma$ eV [8]\\
\hline
$3s3p$   & 17.70 & 0.157   & 17.68 & 0.169  \\
\hline
$3s4p$   & 18.85 & 0.0318  & 18.83 & 0.0321 \\
\hline
$3s5p$   & 19.45 & 0.00601 & 19.41 & 0.0062 \\
\hline
$3s6p$   & 19.73 & 0.0157  & 19.68 & 0.0161 \\
\hline
$3p4s$   & 19.81 & 0.00328 & 19.77 & 0.0033 \\
\hline
$3s7p$   & 19.89 & 0.0274  & 19.82 & 0.0282 \\
\hline
$3s8p$   & 19.95 & 0.0140  & 19.93 & 0.0143 \\
\hline
$^{1}D$  & E, eV & $\Gamma$ eV & E, eV [8]& $\Gamma$ eV [8]\\
\hline
$3s3d$   & 17.62 & 0.0214  & 17.56 & 0.0220 \\
\hline
$3p^{2}$ & 18.31 & 0.0224  & 18.67 & 0.0230 \\
\hline
$3s4d$   & 19.09 & 0.0378  & 19.09 & 0.0389 \\
\hline
$3s5d$   & 19.60 & 0.0121  & 19.56 & 0.0128 \\
\hline
$3d^{2}$ & 19.67 & 0.00789 & 19.63 & 0.0796 \\
\hline
$3s6d$   & 19.81 & 0.00331 & 19.79 & 0.0034 \\
\hline
$^{1}F$  & E, eV & $\Gamma$ eV & E, eV [8]& $\Gamma$ eV [8]\\
\hline
$3p3d$   & 18.96 & 0.0203  & 18.95 & 0.0214 \\
\hline
$3s4f$   & 19.43 & 0.0149  & 19.43 & 0.0155 \\
\hline
$3s5f$   & 19.72 & 0.0070  & 19.70 & 0.00717\\
\hline
$3s6f$   & 19.88 & 0.0023  & 19.85 & 0.00235\\
\hline
$3s7f$   & 19.95 & 0.00021 & 19.94 & 0.00023\\
\hline
$3s8f$   & 19.97 & 0.0019  &   -   &    -   \\ [2mm]
\hline
\end{tabular}

\end{center}

\end{table}

\begin{table}[h]
\noindent\caption{Comparison of the energies obtained with the use of the ICCNR method for the AIS of a beryllium atom, which are located between the corresponding first and second ionization thresholds, with the results of other authors}\tabcolsep4.2pt

\begin{center}

\begin{tabular}{|c|c|c|c|c|}
\hline
\rule{0pt}{5mm} $^{1}P$  & E, eV & E, eV [5]& E, eV [6]& E, eV [7] \\
\hline
$2p3s$ & 10.71 & 10.71  & 10.93 & 10.77 \\
\hline
$2p3d$ & 10.84 & 11.86  & 11.86 & 11.86 \\
\hline
$2p4s$ & 12.03 & 11.97  & 12.10 & 12.07 \\
\hline
$2p4d$ & 12.42 & 12.47  & 12.50 & 12.49 \\
\hline
$^{1}P$& E, eV & E, eV [8]& E, eV [9]& E, eV [10] \\
\hline
$2p3s$ & 10.71 & 10.73  & 10.63 & 10.91 \\
\hline
$2p3d$ & 10.84 & 11.85  & 12.03 & 11.83 \\
\hline
$2p4s$ & 12.03 & 12.09  & 12.09 & 12.09 \\
\hline
$2p4d$ & 12.42 & 12.49  & 12.61 & 12.44 \\ [2mm]
\hline
\end{tabular}

\end{center}

\end{table}

\section{Conclusion}

The method of interacting configurations in the complex number representation, which was applied earlier to the description of quasistationary states of a helium atom [1--3], was used to calculate the ionization processes of more complicated atomic systems. The spectroscopic characteristics of the lowest AIS of the Ве atom were studied in the problem of the electron-impact ionization of these atom. The energies and the widths of the lowest autoionizing states ($^{1}S,\,^{1}P,\,^{1}D,$ and $^{1}F$) of Be atom were calculated. The calculation results were compared with known experimental data and calculations on the basis of other methods. Hence, we may draw conclusion about a successful verification of the method proposed for the calculation of AIS of complex atoms and the processes of electron ionization and excitation of such atoms.




%

\end{document}